# Spin: The Classical to Quantum Connection

James R. Bogan
Department. of Physics
Southern Oregon University
Ashland, OR 97520
boganj@sou.edu

**Abstract**

We show there exists an exact and continuous gauge transformation between the Hamilton-Jacobi equation of classical mechanics, and the time-dependent Schrödinger equation of quantum mechanics. The transformation parameter is spin-dependent, and is a function of the quantum potential of Bohmian mechanics.

**I. Introduction**

In 1926, Schrödinger conjoined DeBroglie's relation between wavelength and kinetic energy, the classical D'Alembert wave equation, and a reasonable ansatz for a stationary wave function, to arrive at the fundamental equation of non-relativistic quantum mechanics [1]. Throughout the remainder of the 20th century, there have been many attempts [2-9] to derive the time-dependent Schrödinger equation (TDSE). Following the approach of Nelson [2], many have employed stochastic methods, which exploit the similarity to the classical diffusion equation. However, physically unimaginable parameters surfaced, such as imaginary diffusion coefficients [4,5] or a universal Brownian motion [6], and at best obtaining only the time-independent Schrödinger equation (TISE). Finally, Frieden [14] has utilized Fisher information theory to deduce the TISE, but failed to derive the TDSE, claiming that it is not possible to derive the TDSE from the Hamilton-Jacobi equation (HJE), even in the limit of Planck's constant vanishing, since the two equations are anisomorphic. On the contrary, we show that the bilinear action gradient in the HJE is coupled to the linear Laplacian operator acting on the wave function through the classical action and wave function modulus. This relation and the inclusion of a parameter to be identified with particle spin, successfully enable the transformation.

A recent book by Namsrai [7] documents attempts to derive the TDSE from stochastic theories, yet as Baublitz [8] points out, all such attempts can be criticized for lack of a physically plausible model for the underlying microscopic fluctuations.

One such avenue which does not suffer from this, is stochastic electrodynamics (SED), which posits a reciprocal exchange of energy between the zero-point fluctuations in vacuum electromagnetic fields, and the stochastic motion of charged elementary particles (*zitterbewegung*)). This process can be shown to reproduce the Planck black-body radiation spectrum, the Bohr radius of the Hydrogen atom, the form of Newtonian gravity, and a physically compelling model for the origin of inertia [9-12], suggesting that zero-point vacuum fluctuations may play a fundamental role on all scales of physical phenomena. An SED-based attempt to derive the TDSE [13] required the *ad hoc* insertion of a `separating function' and Planck's constant at the end of the derivation to maintain dimensional consistency and set the scale of the action, appearing unaesthetic and contrived at best. Indeed, any attempt to derive quantum mechanics from classical mechanics must necessarily introduce Planck's constant as an empirical scale factor. Here, we normalize the classical action to $\hbar$, *ab initio*, so that it functions as a phase angle variable.

In this paper, we show that the Hamilton-Jacobi equation HJE of classical mechanics, and the TDSE of quantum mechanics are mutually derivable. Just as in Scharf's [21] method, Maxwell's equations are derived from Gauss' law, the Lorentz transformations, and the velocity-field relations of special relativity, the TDSE follows from the HJE, a gauge transformation, and a spin-velocity field term from the Pauli current. This irrotational component of the Pauli current is determined by the *zitterbewegung*, which is in turn, coupled through the particle spin, to the quantum potential of Bohmian mechanics [19]. Finally, we argue that this term determines the gauge transformation parameter, and derive a semiclassical expression for the spin angular momentum.

**II. Theory**

To obtain a wave equation governing the propagation of DeBroglie waves associated with an elementary quantum particle, we employ the HJE, the gradient of which is equivalent to Newton's 2nd law [18]. Consider the complex, polar representation of the wave function, given by

(1) $$\Psi(x,t) = R(x,t)e^{iS/\hbar},$$

where we allow the amplitude R to be complex in general, and take S(x,t) to be the real classical action, normalized by Planck's constant to produce a dimensionless phase.

Writing the action as a function of $\Psi$ and R, we have,

(2) $$S(x,t) = \frac{\hbar}{i}(\ln(\Psi/R));$$

The relevant spatial and temporal first and second partial derivatives of S(x,t) are:

(3a) $$\nabla S = \frac{\hbar}{i}(\frac{1}{\Psi}\nabla\Psi - \frac{1}{R}\nabla R)$$

(3b) $$\frac{\partial S}{\partial t} = \frac{\hbar}{i}(\frac{1}{\Psi}\frac{\partial \Psi}{\partial t} - \frac{1}{R}\frac{\partial R}{\partial t})$$

(3c) $$\nabla^2 S = \frac{\hbar}{i}(\frac{1}{\Psi}\nabla^2\Psi - (\frac{1}{\Psi}\nabla\Psi)^2 - \frac{1}{R}\nabla^2 R + (\frac{1}{R}\nabla R)^2)$$

(4) $$(\nabla S)^2 = -\hbar^2((\frac{1}{\Psi}\nabla\Psi)^2 + (\frac{1}{R}\nabla R)^2 - \frac{2}{\Psi R}\nabla\Psi\cdot\nabla R)\,,\text{ which is (3a) squared.}$$

Substituting for the first term in (4) from (3c) and explicitly writing out $\nabla\Psi$ (see eq.(15a)) we obtain

(5) $$(\nabla S)^2 = -\hbar^2(\frac{1}{\Psi}\nabla^2\Psi - \frac{1}{R}\nabla^2 R - \frac{i}{\hbar}\nabla^2 S - \frac{2i}{\hbar R}\nabla R\cdot\nabla S)\,;$$

The HJE [16] is given by,

(6) $$\frac{1}{2m}(\nabla S)^2 + V(x) = -\frac{\partial S}{\partial t}\,;$$

Inserting (5) & (3b) into (6), and multiplying by $\Psi$ leads to,

(7) $$\frac{-\hbar^2}{2m}\nabla^2\Psi + V(x)\Psi + \frac{\hbar}{i}\frac{\partial\Psi}{\partial t} = (\frac{-\hbar^2}{2m}(\nabla^2 R + \frac{iR}{\hbar}\nabla^2 S + \frac{2i}{\hbar}\nabla R\cdot\nabla S) + \frac{\hbar}{i}\frac{\partial R}{\partial t})e^{iS/\hbar})$$

Multiplying (7) by $\dfrac{\Psi^*}{i\hbar}$, we have,

(8) $$\{(\frac{-\hbar^2}{2m}\frac{\partial^2\Psi}{\partial x^2} + V(x)\Psi + \frac{\hbar}{i}\frac{\partial\Psi}{\partial t})\frac{\Psi^*}{i\hbar}\} = (\frac{-\hbar^2}{2m}(\nabla^2 R + \frac{iR}{\hbar}\nabla^2 S + \frac{2i}{\hbar}\nabla R\cdot\nabla S) + \frac{\hbar}{i}\frac{\partial R}{\partial t})\frac{R^*}{i\hbar}$$

Adding (8) to its complex conjugate {c.c.}, results in,

(9) $$\{\frac{\Psi^*}{i\hbar}(-\frac{\hbar^2}{2m}\nabla^2\Psi + V(x)\Psi + \frac{\hbar}{i}\frac{\partial\Psi}{\partial t})\} + \{c.c.\} = \frac{-\hbar}{2mi}(R^*\nabla^2 R - R\nabla^2 R^*) - \frac{1}{m}|R|^2\nabla^2 S$$
$$-\frac{1}{m}\nabla S\cdot\nabla|R|^2 - \frac{\partial}{\partial t}|R|^2$$

Factoring out the first spatial derivative <u>from the right side</u> of (9), and re-writing it as

(10) $$-\nabla\cdot(\frac{\hbar}{2mi}(R^*\nabla R - R\nabla R^*) + \frac{1}{m}|R|^2\nabla S) - \frac{\partial}{\partial t}|R|^2 = c_0\,;$$

Insisting on conservation of probability, forces (10) to vanish, and we obtain the equation of continuity (EOC), with the probability current density given by

(11) $$\vec{J}(x,t) = \frac{\hbar}{2mi}(R^*\nabla R - R\nabla R^*) + \frac{1}{m}|R|^2\nabla S\,;$$

which is easily seen to be the normal form of the current density, given in standard quantum mechanics texts [16,17] alongside the probability density $\rho$ where

(12) $$\vec{J}(x,t) = \frac{\hbar}{2mi}(\Psi^*\nabla\Psi - \Psi\nabla\Psi^*), \text{ and } \rho = |R|^2 = |\Psi|^2 \text{ respectively.}$$

Since the right side of (9) vanishes, we have

(13) $$\{\frac{\Psi^*}{i\hbar}(-\frac{\hbar^2}{2m}\nabla^2\Psi + V(x)\Psi + \frac{\hbar}{i}\frac{\partial\Psi}{\partial t})\} + \{c.c\} = 0$$

Since both { } brackets are the complex conjugates of the other, the only non-trivial way (13) can hold is for the bracketed quantities to vanish individually, which leads of course, to the TDSE,

(14) $$\frac{-\hbar^2}{2m}\nabla^2\Psi + V(x)\Psi = i\hbar\frac{\partial\Psi}{\partial t}$$

Now that we have obtained the TDSE from the HJE, we apply the same procedure to the wave function, and show that the same ansatz will, perhaps not surprisingly, produce the HJE, again accompanied by the equation of continuity for the probability current density. We begin with the TDSE, calculating the first & second derivatives of the wave function (1)

(15a) $$\nabla\Psi = (\nabla R + \frac{iR}{\hbar}\nabla S)e^{iS/\hbar}$$

(15b) $$\frac{\partial\Psi}{\partial t} = (\frac{\partial R}{\partial t} + \frac{iR}{\hbar}\frac{\partial S}{\partial t})e^{iS/\hbar}$$

(15c) $$\nabla^2\Psi = (\nabla^2 R + 2\frac{i}{\hbar}\nabla R\cdot\nabla S + \frac{iR}{\hbar}\nabla^2 S - \frac{R}{\hbar^2}(\nabla S)^2)e^{iS/\hbar};$$

Inserting (15b) & (15c) into the TDSE we obtain,

(16) $$\frac{-\hbar^2}{2m}(\nabla^2 R + 2\frac{i}{\hbar}\nabla R\cdot\nabla S + \frac{iR}{\hbar}\nabla^2 S - \frac{R}{\hbar^2}(\nabla S)^2)e^{iS/\hbar} + V(x)Re^{iS/\hbar} = i\hbar(\frac{\partial R}{\partial t} + \frac{iR}{\hbar}\frac{\partial S}{\partial t})e^{iS/\hbar}$$

Dividing by $\Psi$ we obtain,

(17) $$\frac{-\hbar^2}{2mR}\nabla^2 R - \frac{i\hbar}{mR}\nabla R\cdot\nabla S - \frac{i\hbar}{2m}\nabla^2 S + \frac{1}{2m}(\nabla S)^2 + V(x) = -\frac{\partial S}{\partial t} + \frac{i\hbar}{R}\frac{\partial R}{\partial t}$$

Collecting only the (4th,5th,6th) real terms, thus isolates the HJE (6); The remaining complex terms, are seen to be identical to the right side of (7), and lead directly to the EOC via steps similar to (8-10), thereby justifying our choice of $c_0 = 0$ in (10). The preceding calculations show that the EOC is a dynamical invariant of the transformation between the HJE & the TDSE, and imply the existence of a corresponding symmetry between the equations of motion of classical & quantum mechanics.

### III. Analysis

In section II., we have demonstrated that there exists a continuous and reciprocal transformation between the HJE and the TDSE. Indeed, the emergence of the correct expression for the quantum mechanical probability current alongside these equations depends critically upon the complex nature of R. This can easily be seen by noting that the difference terms in (11) originate solely because R, R*, and their respective derivatives are not equal. This is in contrast to Bohmian mechanics, in which the probability amplitude is taken, *apriori*, to be a real, rather than a complex function. Since R is complex, we again express it in polar representation as,

$$(18) \qquad R = A(x,t)e^{i\theta(x,t)}$$

the amplitude and phase being real-valued. Inserting eq.(18) into (11), gives

$$(19) \qquad \vec{J}(x,t) = \frac{A^2}{m}(\nabla S + \hbar \nabla \vartheta) = \frac{A^2}{m}\nabla \tilde{S}, \quad \text{where we identify } \tilde{S} = S + \hbar \theta$$

as the gauge transformed action, which originates in the implicitly complex nature of R in (1), and is now composed of both classical & quantum components, $S$ and $\hbar \vartheta$ respectively. In what follows, we will argue that $\vartheta$ is associated with the internal kinetic energy in the center-of-mass frame (*zitterbewegung*), which in turn, is identified with the spin of the particle.

The HJE is gauge invariant and may be written as

$$(20) \qquad \frac{1}{2m}(\nabla \tilde{S})^2 + V(x) = -\frac{\partial \tilde{S}}{\partial t}$$

Here, we consider only stationary states, for which the right side of (20) is identical to the total energy, E. Inserting $\tilde{S}$ into (20), and expanding the square, we have

$$(21) \qquad \frac{1}{2m}((\nabla S)^2 + 2\hbar \nabla \vartheta \cdot \nabla S + (\hbar \nabla \vartheta)^2) + V(x) = Ecl + Q,$$

where we have decomposed the energy into classical and quantum components.

Noting that the first term on the left side of (21) is just the classical kinetic energy, we identify

$$(22) \qquad Q = \frac{\hbar}{m} \nabla \vartheta \cdot \nabla S + \frac{\hbar^2}{2m} (\nabla \vartheta)^2$$

In Bohmian mechanics, quantum effects are accounted for in the H-J equation, by the addition of a `quantum potential', Q, added to the sum of the classical kinetic and potential energies, given by

$$(23) \qquad Q = -\frac{\hbar^2}{2m} \frac{\nabla^2 A}{A}$$

Recami & Salesi [19] have shown how the internal (*zitterbewegung*) velocity is connected to spin-1/2, and can be utilized to derive the quantum potential of BM. For the case of spin eigenstates in which the spin vector is constant, the circulation term in the Pauli current goes over to,

$$(24) \qquad \vec{V} = \frac{\nabla \rho \times \vec{s}}{m\rho} = \vec{Vs} \times \hat{s}$$

In a recent extension of their work, Esposito [20] has generalized this result for arbitrary spin, and by only considering the spin direction, shows that the quantum potential can be identified with this internal kinetic energy of the particle.

Comparison of our (19) with Esposito's (31), leads to the relation

$$(25) \qquad \frac{\hbar}{m} \nabla \vartheta = \vec{Vs} \times \hat{s}$$

where, as it is argued in [20], $\vec{Vs}$ is a <u>velocity field</u> associated with & orthogonal to the spin of the particle, and $\hat{s}$ is a unit vector in the direction of the spin. Notice (25) is irrotational; thus one cannot integrate directly to obtain $\vartheta$, since Stoke's theorem causes the cross-product term to vanish as a surface integral. However, Esposito's representation of the quantum potential in terms of the velocity field is given by,

$$(26) \qquad Q = -\frac{1}{2}(m\vec{Vs}^2 + \hbar \nabla \cdot \vec{Vs})$$

and allows one to obtain $\vartheta$.

Solving (25) for the spin velocity field, we have

$$\vec{Vs} = \frac{\hbar}{m}\hat{s} \times \nabla\vartheta \quad (27)$$

Substituting into (26), the divergence term vanishes, leaving

$$Q = \frac{-m}{2}(\frac{\hbar}{m}\hat{s}\times\nabla\vartheta)^2 = -\frac{\hbar^2}{2m}(\nabla\vartheta)^2 \quad (28)$$

This is consistent with the geometrical requirement in [20], that the particle center of mass motion is driven by the classical action gradient, and is <u>orthogonal</u> to the plane containing the spin velocity $\vec{Vs}$ and spin direction, $\hat{s}$, (i.e.,the plane containg the phase gradient). Thus the first term in (22) vanishes, and it becomes equivalent to (28), leading to

$$\nabla^2 A = (\nabla\vartheta)^2 A \quad (29)$$

Now, one may integrate the gradient as,

$$\vartheta = \pm\int(\frac{\nabla^2 A}{A})^{1/2} dx \quad (30)$$

This result is empirically satisfying, since the measurement of a spin projection, results in spin-up and spin-down components, which manifest explicitly here.

Thus in the *zitterbewegung* picture, not only is an intrinsic angular momentum embedded in the Schrödinger equation, as Bohmian mechanics makes evident, but the reciprocal transformation from the HJE to the TDSE derives its' physical origin from spin as well.

Finally, from (25), we see that the Laplacian of $\vartheta$ vanishes. Given that the spin vector breaks the isotropy of the space around the particle, we posit a cylindrical symmetry, and a radial dependence to $\vartheta$. The general solution to such a Laplacian is given by,

$$\vartheta(r) = \alpha\ln(\frac{r}{r_0}); \quad \text{Thus the gradient becomes,} \quad (31)$$

$$\nabla\vartheta = \frac{\alpha}{r}\hat{r} \quad ; \quad \text{Inserting (32) into (27) we obtain} \quad (32)$$

(33) $$\vec{V}s = \frac{\hbar\alpha}{mr}\hat{s}\times\hat{r}$$ Which allows us to express the spin vector as

(34) $$\vec{s} = \hbar\alpha\hat{s} = r\hat{r}\times m\vec{V}s .$$ Or, perhaps more transparently,

(35) $\vec{s} = \vec{r}\times\vec{P}s$ , a semi-classical expression for the intrinsic spin angular

momentum. Notice that (34) allows us to define $\alpha = \frac{|\vec{s}|}{\hbar}$ in (31). Thus $\vartheta$ becomes a gauge of the spin angular momentum. Particularly significant, is that for $r = r_0$ in (31), both $\vartheta$ and the quantum potential vanish, also severing the connexion between classical & quantum mechanics. If one associates the Compton wavelength with the scale parameter $r_0$ in (31), we obtain from (33),

(36) $|V_s| = \alpha c$ , similar to the result of Barut & Zanghi [22].

**Conclusions**

We have shown that there is a simple inference of Schrödinger's time-dependent wave equation, from the Hamilton-Jacobi equation, which provides for an exact and continuous reciprocal transformation between classical and non-relativistic quantum mechanics. The transformation is made possible by a simple gauge transformation representing particle spin, and the reasonable requirement that probability be conserved. This leads to a semi-classical expression for the spin angular momentum.

**Acknowledgments**